\documentclass[aps,prl,reprint,superscriptaddress]{revtex4-2}
\usepackage{graphicx}
\usepackage{amsmath}
\usepackage{mathrsfs}
\usepackage{array}
\usepackage[mathlines]{lineno}
\usepackage[breaklinks=true,colorlinks=true,linkcolor=blue,urlcolor=blue,citecolor=blue]{hyperref}
\graphicspath{{../figures/}}

\begin{document}

\title{Experimental Measurements of Ion Diffusion Coefficients and Heating in a Multi-Ion-Species Plasma Shock} 
\author{F. Chu}
\email[]{fchu@lanl.gov}
\affiliation{Los Alamos National Laboratory, Los Alamos, New Mexico 87545, USA}
\author{A. L. LaJoie}
\affiliation{Department of Electrical and Computer Engineering, University of New Mexico, Albuquerque, New Mexico 87131, USA}
\affiliation{Los Alamos National Laboratory, Los Alamos, New Mexico 87545, USA}
\author{B. D. Keenan}
\affiliation{Los Alamos National Laboratory, Los Alamos, New Mexico 87545, USA}
\author{L. Webster}
\affiliation{Department of Electrical and Computer Engineering, University of New Mexico, Albuquerque, New Mexico 87131, USA}
\author{S. J. Langendorf}
\affiliation{Los Alamos National Laboratory, Los Alamos, New Mexico 87545, USA}
\author{M. A. Gilmore}
\affiliation{Department of Electrical and Computer Engineering, University of New Mexico, Albuquerque, New Mexico 87131, USA}
\date{\today}

\begin{abstract}
Collisional plasma shocks generated from supersonic flows are an important feature in many astrophysical and laboratory high-energy-density plasmas. Compared to single-ion-species plasma shocks, plasma shock fronts with multiple ion species contain additional structure, including interspecies ion separation driven by gradients in species concentration, temperature, pressure, and electric potential. We present time-resolved density and temperature measurements of two ion species in collisional plasma shocks produced by head-on merging of supersonic plasma jets, allowing determination of the ion diffusion coefficients. Our results provide the first experimental validation of the fundamental inter-ion-species transport theory. The temperature separation, a higher-order effect reported here, is valuable for advancements in modeling HED and ICF experiments.
\end{abstract}


\maketitle

\textit{Introduction}.---Shocks generated from supersonic flows are abundant in a variety of plasma environments, including astrophysical systems \cite{ryu_cosmological_2003, johlander_rippled_2016} and high-energy-density (HED) laboratory experiments \cite{remington_high_2005, rochau_radiating_2008, drake_perspectives_2009, rinderknecht_highly_2018}. Compared to hydrodynamic shocks, plasma shocks may be governed primarily by Coulomb interactions between charged particles (collisional plasma shocks) \cite{hough_electron_2009} or by the influence of the electromagnetic fields (collisionless shocks) \cite{treumann_fundamentals_2009}. In addition, kinetic effects occurring in plasma shocks usually cannot be fully described using hydrodynamic shock theory \cite{rosenberg_exploration_2014, kagan_self-similar_2015, simakov_plasma_2017, zhang_species_2020}. When multiple ion species are present in a collisional plasma shock, additional features can be introduced to the shock front structure, one of which is interspecies ion separation resulting from differential gradient-driven ion diffusion \cite{amendt_plasma_2010, bellei_species_2013, vold_multi-species_2019}.

Work on understanding the ion diffusion mechanisms in multi-ion-species plasma shocks began only within the last decade \cite{rygg_tests_2006}. Ever since, many theoretical, computational, and experimental efforts have been made to study inter-ion-species diffusion in the context of inertial confinement fusion (ICF) experiments, in which mixtures of deuterium and tritium (DT) are compressed by strong shock waves to achieve fusion reactions \cite{kagan_electro-diffusion_2012, kagan_thermo-diffusion_2014, simakov_electron_2014, molvig_classical_2014, rinderknecht_ion_2015, simakov_hydrodynamic_2016, simakov_hydrodynamic_2016-1, hsu_observation_2016, kagan_influence_2017, keenan_ion_2018, rinderknecht_measurements_2018, joshi_progress_2019, byvank_observation_2020, mohammed_ion_2022}. Recently, inter-ion-species diffusion theory has also been adopted in magnetized liner inertial fusion (MagLIF) to investigate energy and magnetic flux losses in a hot magnetized plasma confined by a cold liner wall \cite{garcia-rubio_mass_2018}. However, the underlying ion diffusion model used in these prior theoretical and simulation works has never been directly validated by experiment. In addition, our understanding of the kinetic signatures that cannot be captured by fluid theory or hydrodynamic simulations, such as non-local- thermodynamic-equilibrium (non-LTE) effects, still remains limited to this day.

In this Letter, we report detailed experimental investigation on the time evolution of the density and temperature of two separate ion species (ArII and NII) in a collisional plasma shock. Our measurements of ion diffusion coefficients, for the first time, not only enable direct validation of the fundamental inter-ion-species transport theory, but can also be used to benchmark ion diffusion models in fusion and space plasma simulations. Furthermore, the temperature separation, a higher-order effect demonstrated in the experiment, provides new data valuable for advancements in modeling HED and ICF experiments.

\textit{Inter-Ion-Species Diffusion}.---In a multi-ion-species plasma, the relationship between the bulk fluid velocity $\textbf{u}$ of the plasma and the flow velocity $\textbf{u}_\alpha$ of each ion species $\alpha$ can be written as $\rho\textbf{u}=\sum_{\alpha} \rho _{\alpha }\textbf{u}_{\alpha }$, where $\rho =\sum_{\alpha }\rho _{\alpha }$ is the total ion mass density and $\rho _{\alpha }=m_{\alpha }n_{\alpha }$ is the partial ion mass density of the species $\alpha$ ($m_{\alpha }$ and $n_{\alpha }$ are the mass and number density of the ion species $\alpha$, respectively) \cite{kagan_thermo-diffusion_2014}. Therefore, the diffusive ion mass flux of each species $\alpha$ in the center-of-mass frame of the plasma is given by $\textbf{i}_{\alpha } = \rho _{\alpha } (\textbf{u}_{\alpha }-\textbf{u})$, where the total mass flux $\sum_{\alpha }\textbf{i}_{\alpha }=0$.


Applying mass conservation to species $\alpha$ and all ions, the flux $\textbf{i}_{\alpha }$ is found to govern the mass concentration $c_\alpha$ of the species $\alpha$ through the continuity equation
\begin{equation}
\label{eq:con}
\rho \frac{\partial c_\alpha }{\partial t}+\rho \textbf{u}\cdot \nabla c_{\alpha }+\nabla \cdot \textbf{i}_{\alpha }=0,
\end{equation} 
where $c_\alpha = \rho _{\alpha }/\rho$. For a system containing only two ion species and close to local thermodynamic equilibrium (LTE), it is shown that the resulting flux has a linear relation with the thermodynamic forces \cite{kagan_thermo-diffusion_2014, simakov_plasma_2017, byvank_observation_2020}. The flux for the lighter ion species takes the form
\begin{eqnarray}
\label{eq:flux}
\textbf{i}=-\rho D &\bigg (& \nabla c+\frac{\kappa _{p}}{P_i}\nabla P_i +\frac{\kappa _{T_i}}{T_i}\nabla T_i \nonumber \\ 
&&+\frac{\kappa _{T_e}}{T_e}\nabla T_e+\frac{e\kappa _{E}}{T_i}\nabla \Phi \bigg ),
\end{eqnarray}
with $e$ the electron charge. The first term represents the classical diffusion flux resulting from the mass concentration gradient, and $D$ is the classical diffusion coefficient. The remaining terms describe the baro-diffusion flux, ion and electron thermo-diffusion flux, and electro-diffusion flux, driven by the respective gradients in total ion pressure $P_i$, ion temperature $T_i$, electron temperature $T_e$, and electrical potential $\Phi$. The analytical expressions for the diffusion coefficients  $D$, $\kappa _{p}$, $\kappa _{T_i}$, $\kappa _{T_e}$, and $\kappa _{E}$ can be found in the Supplemental Material.

At the shock boundary, Eq.~(\ref{eq:flux}) predicts that all the diffusion flux terms for the lighter ion species point from the post-shock region to the pre-shock region except the relatively small electron thermo-diffusion flux, resulting in a net flux pointing in the same direction. Therefore, to satisfy the condition $\sum_{\alpha }\textbf{i}_{\alpha }=0$, the diffusion flux for the heavier ion species has to point in the opposite direction in the center-of-mass frame, causing the ion species to separate in the plasma shock \cite{byvank_observation_2020}.

\textit{Experiment}.---To experimentally investigate ion diffusion within a shock front, we create multi-ion-species plasma shocks by colliding two plasma jets head-on in a cylindrical vacuum chamber of 137 cm length and 76 cm diameter. The jets are produced by capacitor-driven plasma guns that accelerate plasma via $j \times B$ force \cite{hsu_experiment_2018, thio_plasma-jet-driven_2019}. At the time when the plasma guns are fired, a gas puff consisting of 50\% Ar and 50\% $\textup{N}_2$ by volume (58.8\% Ar and 41.2\% $\textup{N}_2$ by mass concentration) is first pre-ionized in the gun nozzle and then accelerated into the vacuum chamber to a speed of $v_\textup{jet}=18$ km/s. Each individual jet collides with an electron and ion temperature $T_e \approx T_i \approx 2$ eV, density $n_e \approx n_i \approx 5 \times 10^{14}$ $\textup{cm}^{-3}$, and mean charge $\bar{Z} \approx 1$ \cite{byvank_observation_2020}. The Mach number is thus $\sim 8.2$ for Ar and $\sim 4.9$ for N in the pre-shock region.

\begin{figure}
\begin{center}
\includegraphics[width=3.37in]{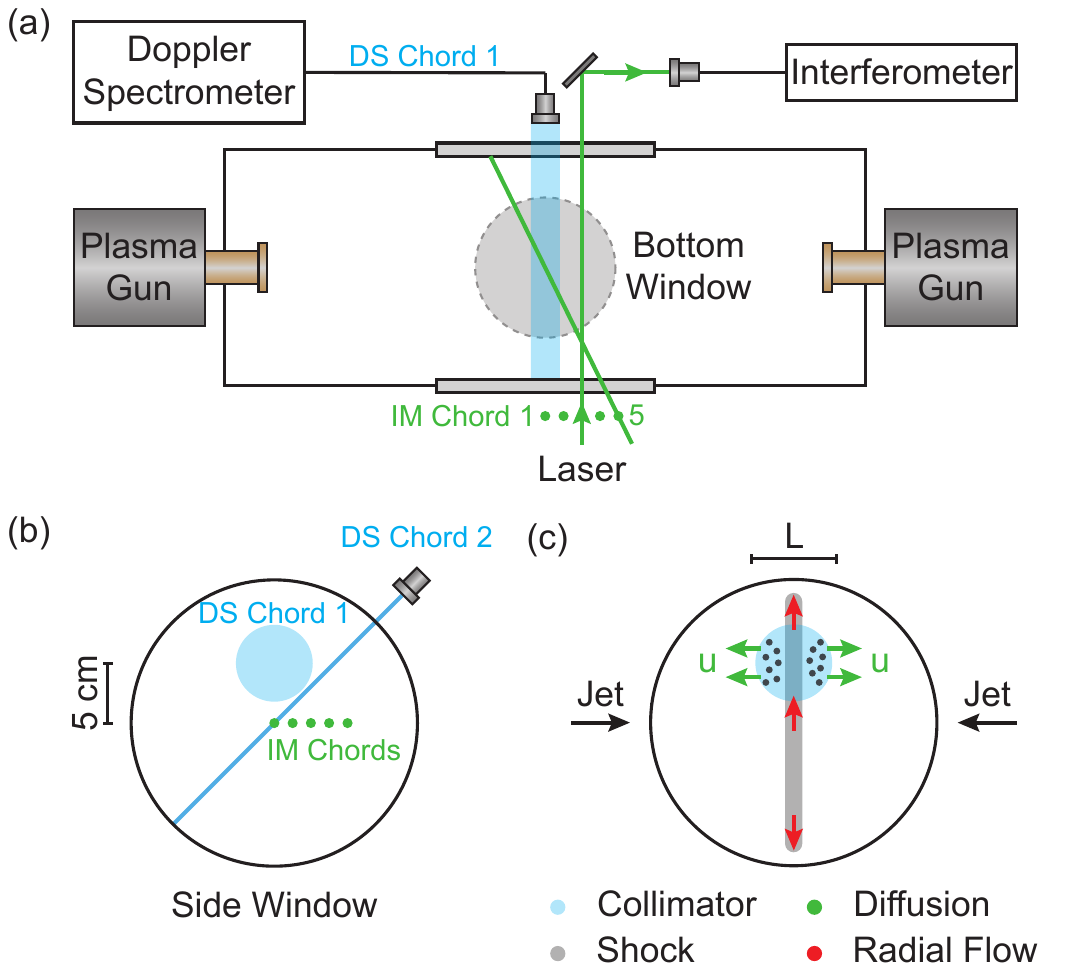}
\caption{Top view (a) and side view (b) of the experimental setup. Plasma jets are generated by two plasma guns and head-on collide in the center of the chamber. The main diagnostics (not drawn to scale) include a two-chord Doppler spectrometer (DS, colored in blue), a five-chord heterodyne interferometer (IM, colored in green), a diagnostic spectrometer, and a fast intensified charge-coupled-device (ICCD) camera. (c) Schematic of shock formation demonstrating that ion diffusion in the axial direction is the only major mechanism that can change the ion density in the collimator viewing volume. The diameter of the collimator $L$=2 cm.}
\label{fig:exp}
\end{center}
\end{figure}

The top view of the experimental setup is depicted in Fig.~\ref{fig:exp}(a). A five-chord heterodyne interferometer is used to measure the time-resolved line-integrated electron density in the plasma shock \cite{merritt_multi-chord_2012, merritt_multi-chord_2012-1}. The laser emission is produced using a 320 mW, 561 nm solid state laser, which is divided into multiple lower-power beams and injected into the chamber through five single-mode fiber optic cables. Each probe beam is $\sim 0.3$ cm in diameter and placed horizontally in the midplane of the chamber at 0, 2, 4, 6, 8 cm with respect to the center. The first four probe beams are directed into the chamber perpendicularly to the side window (parallel with the shock front) while the fifth at an angle of $75^{\circ}$. These chord angles are chosen so that we can use chord 5 to measure the average electron density in the plasma shock and chords 1 -- 4 to determine the bulk flow velocity in the post-shock region via time-of-flight analysis. After passing through the plasma, the probe beams are compared with a reference beam of the interferometer in the Mach-Zender configuration, and the line-integrated electron density is computed based on the relative phase shift \cite{hutchinson_principles_2002}.

\begin{figure}
\begin{center}
\includegraphics[width=3.37in]{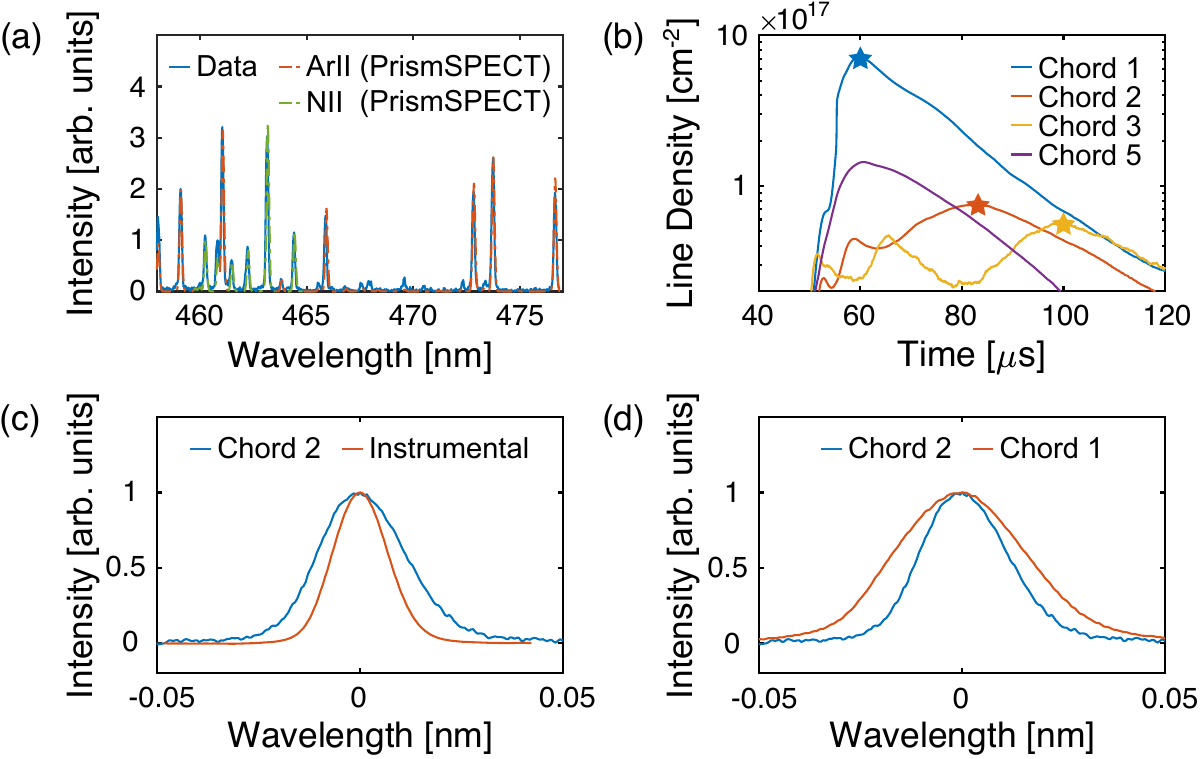}
\caption{(a) Example of broadband ion emission spectrum from the diagnostic spectrometer, along with the calculated spectrum using PrismSPECT at $T_e \approx 2.2$ eV. (b) Example of line-integrated electron density from interferometry. The peak line densities of chord 1, 2, and 3 are marked by stars, which are used in time-of-flight analysis to determine the plasma bulk flow velocity in post-shock region. (c) Example of Doppler-broadened ArII 434.9 nm line. The orange curve shows the instrumental broadening of the same line without the Doppler effects. (d) Comparison between Doppler-broadened ArII 434.9 nm lines obtained from DS chord 1 and 2. The center of profiles in (c) and (d) is shifted to 0 nm.}
\label{fig:id}
\end{center}
\end{figure}

As the density in our plasma shock is not high enough for Stark broadening to be appreciable \cite{purcell_stark_1984, nick_experimental_1986, langendorf_experimental_2018, blagojevic_stark_2021}, Doppler broadening caused by ion thermal motion becomes the dominant source of spectral line broadening in the plasma \cite{langendorf_experimental_2018}. A high-resolution Doppler spectrometer (DS) is utilized to measure the ion temperature in the shock through examining the width of the Doppler-broadened ion lines. The light emission from the plasma is collected using two collimators of $L=2$ cm in diameter. Chord 1 is aligned with the central vertical plane of the chamber and held at 5 cm above the horizontal midplane, whereas chord 2 is in the vertical midplane but at an angle of $\sim 45^{\circ}$ with respect to the horizontal plane, as shown in Fig.~\ref{fig:exp}(b). The light collected is then fielded by a double-pass high-resolution monochromator (McPherson 2062DP) with 2400 $\textup{mm}^{-1}$ grating, and the resulting spectrum is captured using a charge-coupled-device (CCD) camera (PCO Pixelfly). The spectral resolving power is $\sim 1 \times 10^5$ at the typical visible wavelengths of interest. In the experiment, we find that the width of the ion Doppler profile obtained from chord 1 is about 1.5 times as that obtained from chord 2 due to the plasma expansion in the radial direction, as shown in Fig.~\ref{fig:id}(d). Therefore, the ion temperature in our plasma shock is estimated based on the measured Doppler broadening of ArII 434.9 nm and NII 399.6 nm lines of chord 2, which removes contributions to the broadening from the post-shock plasma radial expansion.

\begin{figure}
\begin{center}
\includegraphics[width=3.37in]{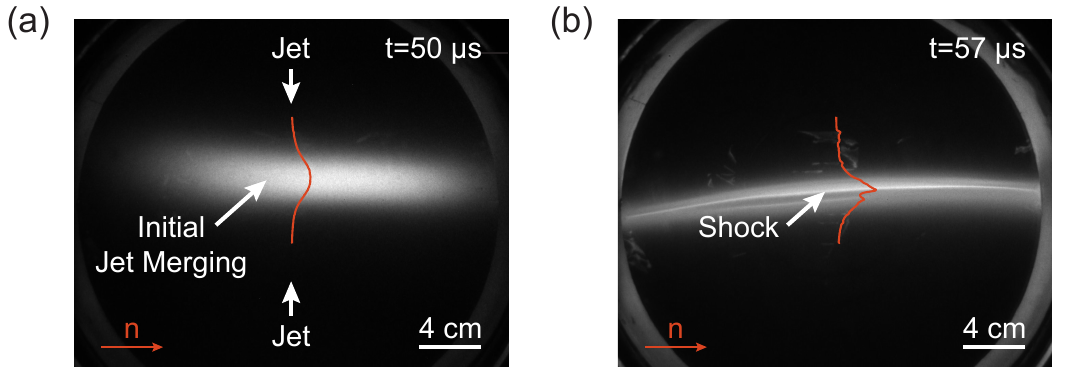}
\caption{(a)--(b) ICCD camera images of shock formation at 50 and 57~$\mu$s, respectively, relative to the moment when the capacitor banks in the guns discharge. The orange line represents the plasma density profile in the center of each image.}
\label{fig:shock}
\end{center}
\end{figure}

The ion emission spectrum is also used to infer the time-resolved density of each ion species in the plasma shock. For an atomic transition from an excited energy state to the lower state, the intensity of the emitted photons is proportional to the density of the excited state, a function strongly dependent on plasma parameters, such as the electron density and temperature \cite{fantz_basics_2006}. By applying the collisional radiative (CR) model to our plasma in PrismSPECT simulation, we can infer the relative density of each ion species from the intensity of the Doppler-broadened spectral lines (see the Supplemental Material for details).

Other diagnostics used in the experiment include a photodiode array at each gun nozzle to measure the jet velocity, a visible diagnostic spectrometer to interpret the electron temperature, and a fast intensified-CCD camera (PCO DiCam Pro) to image the shock formation. The diagnostic spectrometer records the broadband ion emission spectrum, which is then compared to PrismSPECT non-LTE modeling to infer the electron temperature in the plasma shock \cite{hsu_experimental_2012, merritt_experimental_2013, langendorf_experimental_2018, byvank_observation_2020}.

\textit{Results}.---Figures~\ref{fig:shock}(a) and \ref{fig:shock}(b) show the time evolution of multi-ion-species plasma jet merging captured by the ICCD camera, referenced to the moment when the capacitor banks in the guns discharge. At 50 $\mu$s, the two jets are seen to reach the center of the chamber and about to collide. A sharp plasma shock front forms in several microseconds at 57 $\mu$s. Figure~\ref{fig:id}(a) shows an example of the broadband ion emission spectrum from the diagnostic spectrometer and a comparison to the PrismSPECT atomic modeling that we use to infer the electron temperature. Figure~\ref{fig:id}(c) shows an example of the Doppler broadened line, where the ion temperature is computed from performing deconvolution of instrumental broadening with the measured spectral profile. The experimentally inferred plasma parameters in the post-shock region are, electron temperature $T_e \approx 2.2$ eV, density $n_e \approx n_i \approx 1 \times 10^{15}$ $\textup{cm}^{-3}$, Knudsen number $\sim 0.4$, peak Ar temperature $T_\textup{\textit{i}-Ar} \approx 17$ eV, and peak N temperature $T_\textup{\textit{i}-N} \approx 9.3$ eV.

\begin{figure}
\begin{center}
\includegraphics[width=3.37in]{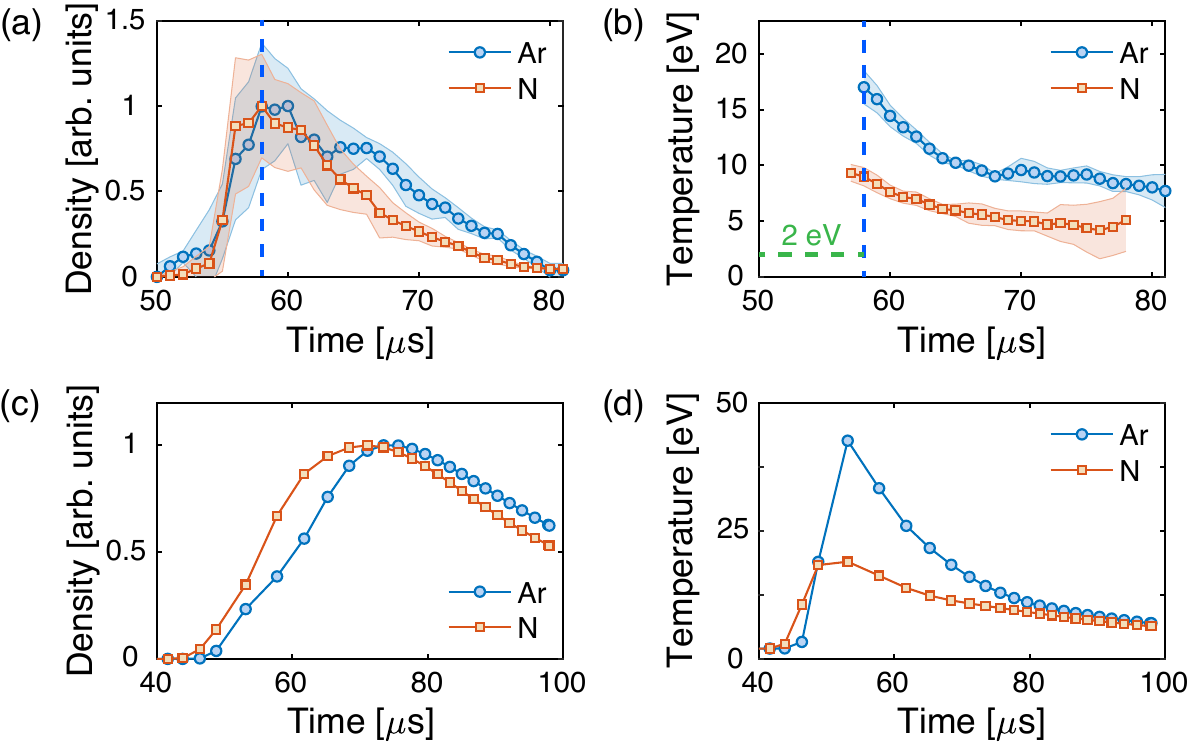}
\caption{(a) Measured ion density from DS chord 1 as a function of time. The density is normalized by the peak value in each curve and the blue vertical dashed line separates the pre- and post-shock region. (b) Measured ion temperature from DS chord 2 as a function of time, demonstrating ion heating and temperature separation. The green horizontal dashed line represents ion temperature in the pre-shock region. Due to shot-to-shot variation in the experiment, each data point in (a) and (b) is averaged over six measurements. The color bands indicate one-standard-deviation uncertainties. (c) Simulated ion density as a function of time using the iFP code. The density is normalized by the peak value in each curve. (d) Simulated ion temperature as a function of time.}
\label{fig:edt}
\end{center}
\end{figure}

One important finding of this experiment is the observations of species separation caused by differential gradient-driven ion diffusion. Figure~\ref{fig:edt}(a) shows the normalized density of Ar and N by their maximum values as a function of time, where the pre- and post-shock region are separated by a blue dashed line. The density of both ion species increases at the same rate within errors at early time; however, at later time as the shock forms, the density of N ions decreases faster than that of Ar ions. This finding suggests that the lighter ion species (N) diffuse faster in the plasma shock gradients, consistent with the theoretical predictions. To bound the ion diffusion velocity, we consider the ion flow caused by plasma expansion and diffusion at the boundary of the collimator viewing volume shown in Fig.~\ref{fig:exp}(c). In the radial direction, the ions enter and leave the volume, resulting in no net change of the ion density. Hence, the ion diffusion in the axial direction becomes the only major mechanism that decreases the ion density in the viewing volume. Solving the continuity equation at the boundary of the viewing volume gives the diffusion velocity as $u=-dn_i/dt \cdot L/2n_i$, where the length scale $L$ is the diameter of the collimator. By inserting the ion density profile in Fig.~\ref{fig:edt}(a), we find that the average diffusion velocity (over $\sim 7$ $\mu$s after the shock forms) for Ar and N is $u_\textup{Ar} = 0.5 \pm 0.1$ km/s and $u_\textup{N} = 1.0 \pm 0.1$ km/s, respectively. We also estimate the plasma bulk flow velocity in the post-shock region $u=\sum_{\alpha }c_{\alpha }u_{\alpha } = 0.7 \pm 0.1$ km/s, relatively consistent with the velocity of $1.1 \pm 0.2$ km/s determined using time-of-flight analysis from the interferometry chords shown in Fig.~\ref{fig:id}(b). Thus, the diffusion velocity for Ar and N in the center-of-mass frame is $v_\textup{Ar}=u_\textup{Ar}-u \approx -190$ m/s and $v_\textup{N} = u_\textup{N}-u \approx 270$ m/s, with the relative diffusion velocity $\Delta v = v_\textup{N}-v_\textup{Ar} \approx 460$ m/s.

\begin{table}[b]
\caption{Comparison of the dominant diffusion coefficients between inter-ion-species diffusion theory and experiment.}
\label{tab:velocity}
\begin{ruledtabular}
\begin{tabular}{lccc}
Diffusion Coefficients [$\textup{m}^2/\textup{s}$] & $D$ & $D \kappa_p$ & $D \kappa_{T_i}$\\
\hline
Theory &4.12 & 1.15 & 1.15\\
Experiment & -- & $0.72 \pm 0.28$ & $0.72 \pm 0.28$\\
\end{tabular}
\end{ruledtabular}
\end{table}



We evaluate the various differential gradient terms in Eq.~(\ref{eq:flux}) to further understand their effects on inter-ion-species diffusion. The gradient of quantity $Q$ is computed as $\nabla Q/Q \approx 2(Q_2-Q_1)/L(Q_2+Q_1)$, where the subscript ``1'' and ``2'' represent the pre- and post-shock region, respectively. The diffusion coefficients, calculated either analytically or numerically (see the Supplemental Material for details), are $D \approx 4.12 $ $\textup{m}^{2}/\textup{s}$, $\kappa_p \approx 0.28$, $\kappa_{T_i} \approx 0.28$, $\kappa_{T_e} \approx -0.06$, and $\kappa_E \approx 0.28$. Since the ambipolar electric field in Eq.~(\ref{eq:flux}) can be estimated as $\nabla \Phi=\nabla T_e/e$ \cite{tang_reduced_2014}, the contributions of each flux term to the diffusion velocity of N ions in the center-of-mass frame are $-40$ m/s for classical diffusion, 233 m/s for baro-diffusion, 199 m/s for ion thermo-diffusion, $-3$ m/s for electron thermo-diffusion, and 4 m/s for electro-diffusion. The above calculations suggest that both baro- and ion thermo-diffusions dominate over other diffusion mechanisms, comparable to the mechanisms that drive species separation in ICF experiments \cite{kagan_thermo-diffusion_2014}. By reasonably assuming $\kappa_p \approx \kappa_{T_i}$ and neglecting other small terms, we are able to experimentally constrain the leading diffusion coefficients $D \kappa_p$ and $D \kappa_{T_i}$ from Eq.~(\ref{eq:flux}). This key experimental result and the comparison to inter-ion-species diffusion theory are listed in Table~\ref{tab:velocity}.

We also experimentally measure heating of the distinct ion species. Figure~\ref{fig:edt}(b) shows the time evolution of Ar and N ion temperature inferred from Doppler spectroscopy. Substantial ion heating and temperature separation are observed during the shock formation, with the temperature elevated from 2 eV to $\sim 17$ eV for Ar and to $\sim 9$ eV for N. Since the ion thermal energy is mainly transferred from its own kinetic energy, a ratio of $\sim 2$ in thermal energy gain is found between Ar and N as a result of their mass ratio ($\sim 2.9$). The discrepancy in these two ratios indicates that some energy loss and equilibration occur in the early time of the shock formation. Subsequent decrease of the ion temperature is observed due to classical thermal equilibration \cite{langendorf_experimental_2018} among both ion species and electrons, with electrons radiatively cooling the plasma in the optically thin conditions.

A 1D Eulerian Vlasov-Fokker-Planck simulation is performed using the iFP code \cite{keenan_shock-driven_2020, taitano_eulerian_2021} (see the Supplemental Material for details), as shown in Figs.~\ref{fig:edt}(c)--\ref{fig:edt}(d), to compare with the experimental results. The simulation in general agrees with the experimental data, successfully predicting a faster density decrease of N, ion heating, and temperature separation in the post-shock region shown in Figs.~\ref{fig:edt}(a)--\ref{fig:edt}(b). One of the noticeable discrepancies between the simulation and experiment is that the simulation predicts a higher ion temperature jump than the data. In the simulation, leading edge ions with a faster velocity reach the center of the chamber first and are heated to a high temperature, as shown in Fig.~\ref{fig:edt}(d) at $\sim 50$ $\mu$s. This effect can be easily seen in the simulation, but difficult to be observed in the experiment because of the low density of this ion group (Fig.~\ref{fig:edt}(c) at $\sim 50$ $\mu$s). In addition, the 1D simulation does not admit the possibility of radial expansion of the post-shock plasma, which is clearly seen in the experiment. This effect provides another mechanism by which kinetic energy is lost from the post-shock region, and contributes to the lower temperatures observed experimentally. It should be noted that the ion temperature separation demonstrated in Figs.~\ref{fig:edt}(b) and \ref{fig:edt}(d) is a higher-order effect in deviations from LTE, whereas the diffusion flux theory assumes that all ion temperatures are equal \cite{simakov_plasma_2017}. Analyzing the simulation output, the average diffusion velocity for N and Ar in the center-of-mass frame is $v_\textup{N} = 300$ m/s and $v_\textup{Ar} = -260$ m/s, respectively. The total relative velocity is thus $\Delta v=560$ m/s, consistent with the experimental results of $\Delta v \approx 460$ m/s. 

\textit{Conclusions}.---We present detailed experimental study of the density and temperature evolution of Ar and N in collisional multi-ion-species plasma shocks. With measurements that resolve the ion diffusion coefficients, we demonstrate for the first time that the dominant diffusion coefficients calculated from fundamental inter-ion-species diffusion theory are in agreement with our experimental results. Furthermore, we find that significant ion heating occurs on the separate ion species during the formation of the shocks, with a temperature jump of $\sim 15$ eV for Ar and $\sim 7$ eV for N, similar to the initial distribution of jet kinetic energy but not simply proportional. We compare our experimental results with 1D Vlasov-Fokker-Planck simulation, and find that in general the calculation reproduces the experimental data, with differences that can be understood from the 3D geometry of the experimental plasma jets and shocks. This investigation on inter-ion-species diffusion in collisional plasma shocks provides useful new data to further understanding of multi-species plasma shocks and their relevance to HED/ICF configurations.


\begin{acknowledgments}

The authors would like to acknowledge T. Byvank, S. D. Baalrud, and J. Dunn for valuable conversations and technical support. Research presented in this Letter was supported by the Laboratory Directed Research and Development program of Los Alamos National Laboratory under Project No. 20200564ECR. The plasma guns used in this work were designed and built by HyperJet Fusion Corporation under funding support of the Advanced Research Projects Agency–Energy (ARPA-E) of the U.S. Department of Energy (DOE) under Contract No. DE-AC5206NA25396 and cooperative agreement No. DE-AR0000566. 



\end{acknowledgments}

\bibliography{../refs}

\end{document}